\documentclass[10pt,twocolumn,letterpaper]{article}

\usepackage[pagenumbers]{cvpr} % To force page numbers, e.g. for an arXiv version
\usepackage{bm}
\usepackage{multirow}
\usepackage{booktabs}

\usepackage[dvipsnames]{xcolor}

\definecolor{cvprblue}{rgb}{0.21,0.49,0.74}
\usepackage[pagebackref,breaklinks,colorlinks,citecolor=cvprblue]{hyperref}

\def\paperID{4796} % *** Enter the Paper ID here
\def\confName{CVPR}
\def\confYear{2024}

\title{DanceMeld: Unraveling Dance Phrases with Hierarchical Latent Codes for Music-to-Dance Synthesis}

\author{Xin Gao, Li Hu, Peng Zhang, Bang Zhang, Liefeng Bo\\
Institute for Intelligent Computing, Alibaba Group\\
\tt\small \{zimu.gx, hooks.hl, futian.zp, zhangbang.zb, liefeng.bo\}@alibaba-inc.com\\
\small \url{https://humanaigc.github.io/dance-meld/}
}

\begin{document}

\twocolumn[{%
\renewcommand\twocolumn[1][]{#1}%
\maketitle
\begin{center}
    \centering
    \captionsetup{type=figure}
    \includegraphics[width=0.9\textwidth]{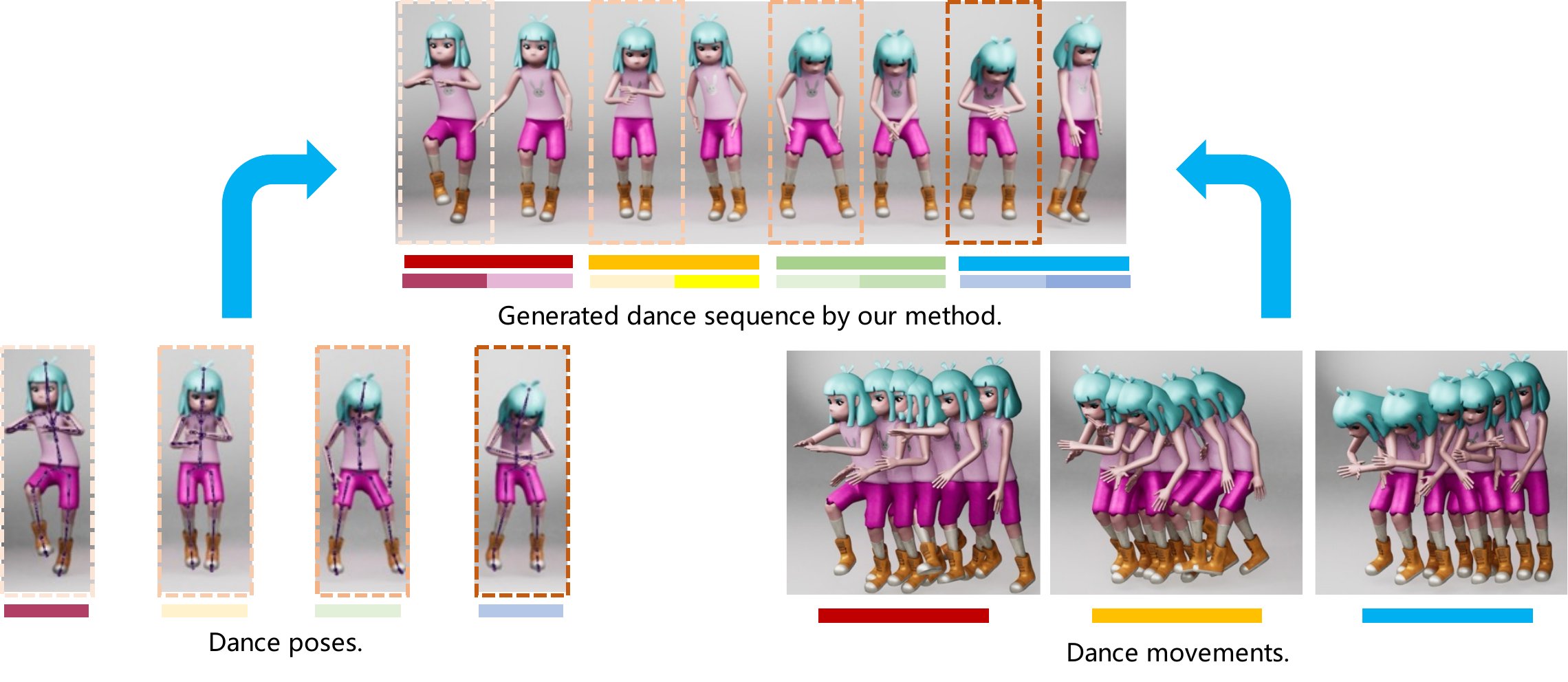}\vspace{-0.2cm}
    \captionof{figure}{In the field of choreography, dance poses composed of a series of basic meaningful body postures, while dance movements can reflect trends, rhythm and energy of the motion. Our method uses a hierarchical VQ-VAE to decouple dance poses and dance movements by representing them with bottom code and top code (short line for bottom code and long line for top code).}

\end{center}%
}]
\maketitle
\begin{abstract}
In the realm of 3D digital human applications, music-to-dance presents a challenging task. Given the one-to-many relationship between music and dance, previous methods have been limited in their approach, relying solely on matching and generating corresponding dance movements based on music rhythm. 
In the professional field of choreography, a dance phrase consists of several dance poses and dance movements. Dance poses composed of a series of basic meaningful body postures, while dance movements can reflect dynamic changes such as the rhythm, melody, and style of dance. 
Taking inspiration from these concepts, we introduce an innovative dance generation pipeline called DanceMeld, which comprising two stages, i.e., the dance decouple stage and the dance generation stage. 
In the decouple stage, a hierarchical VQ-VAE is used to disentangle dance poses and dance movements in different feature space levels, where the bottom code represents dance poses, and the top code represents dance movements. 
In the generation stage, we utilize a diffusion model as a prior to model the distribution and generate latent codes conditioned on music features. 
We have experimentally demonstrated the representational capabilities of top code and bottom code, enabling the explicit decoupling expression of dance poses and dance movements. 
This disentanglement not only provides control over motion details, styles, and rhythm but also facilitates applications such as dance style transfer and dance unit editing.
Our approach has undergone qualitative and quantitative experiments on the AIST++ dataset, demonstrating its superiority over other methods. 
%The relevant codes will be open sourced.
\end{abstract}

\section{Introduction}
\label{sec:intro}

Virtual characters have been popular in various domains, including virtual idols, virtual customer service, and virtual assistants. Driving virtual characters to dance with music is an important application within the gaming industry, film production and AR/VR experiences. However, creating dance animations demands professional expertise to match intricate movements with diverse music styles, rhythm, and melodies. As a result, automated dance generation holds substantial promise, offering a straightforward and cost-effective means for individuals to produce and edit dance sequences for 3D characters, reducing the need for extensive manual effort during the production process.
Music-driven dance generation remains an ongoing research challenge owing to the intricate one-to-many relationships that exist between music and dance.
Recent studies have aimed to address this challenge using machine learning techniques such as Generative Adversarial Networks (GANs) \cite{li2022danceformer, lee2019dancing}, autoregressive models \cite{kim2022brand, li2022danceformer, sun2022you, li2021ai, ye2020choreonet, huang2020dance}, and diffusion models \cite{dabral2023mofusion, tseng2023edge, li2023finedance, alexanderson2023listen}. Unfortunately, most previous music-to-dance methods treat the task as a general human motion generation task, mapping frame-level dance motion directly to sequential music, without taking choreography knowledge into account.

% However, these approaches often struggle to produce consistent and accurate dance sequences, leading to strange poses and uncoordinated movements. 
To tackle this issue, we refer to expert knowledge of professional choreography \cite{blom1982intimate, minton2017choreography}. Dance phrase, which expresses the complete meaning of a smallest dance sequence, serves as the basic unit in choreography. To create a dance phrase, the choreographer should: 1) conceive several meaningful dance poses, 2) connect these dance poses with appropriate dance movements in terms of space, shape, time, energy and emotion from music. Therefore, dance poses and dance movements are two key elements in professional choreography. 
Despite drawing inspiration from the knowledge mentioned above, it is still challenge to model dance poses and dance movements. In real dance motions, poses and movements do not exhibit a straightforward one-to-one mapping, they are intricately intertwined, which represent artistic combinations that are influenced by the dance genre, style, and the rhythm and melody of the music, making it difficult to disentangle. Recently, Bailando \cite{siyao2022bailando} proposes to establish a dancing pose unit dictionary to learn a dancing-style subspace. However, it still needs an extra actor-critic learning with a reward function to align between motion tempo and music beats.

In this paper, we introduce a novel dance generation pipeline called DanceMeld that comprising two stages: the dance decouple stage and the dance generation stage. 
In the dance decouple stage, inspired by previous works in text-to-image synthesis, we employ a hierarchical Vector Quantized Variational Autoencoder (VQ-VAE) \cite{van2017neural, razavi2019generating} to model dance poses and dance movements separately in different feature space levels, where the bottom latent code aims to indicates a certain reasonable dance poses, and the top latent code energizes the representation of movement process which capturing aspects such as orientation, tendency, speed and beat of the motion. 
In the dance generation stage, we leverage a diffusion model as a prior to generate the sequence of latent codes based on the given music. These predicted code sequences are decoded into dance sequences using a motion decoder. 
In summary, our contributions are as follows:
\begin{itemize}
    \item We propose DanceMeld, a two-stage music-to-dance framework that decouple dance poses and dance movements by representing them with bottom code and top code by a hierarchical VQ-VAE, experiments demonstrate the interpretability of the latent codes.
    \item A diffusion model is used as a prior to represent the distribution of latent codes and generate dance sequences conditioned on music, which enables various applications such as dance style transfer and dance unit editing.
    % \item By incorporating domain-specific expertise from choreography, we devise a hierarchical VQ-VAE to decouple dance pose and dance movement by representing them with bottom code and top code, respectively. During the motion generation phase, the diffusion model is employed as the prior to model the distribution of latent codes.
    % \item We have demonstrated the interpretability of bottom code and top code through experiments. This decoupled design enables us to control the generation of dance motions. For example, dance style transfer, dance units editing and other applications can be achieved.
    \item Our approach has been validated both qualitatively and quantitatively on the AIST++ dataset compared to the state-of-the-art methods, demonstrating its effectiveness and superiority.
\end{itemize}

\begin{figure*}[htp]
    \centering
    \includegraphics[width=1\linewidth]{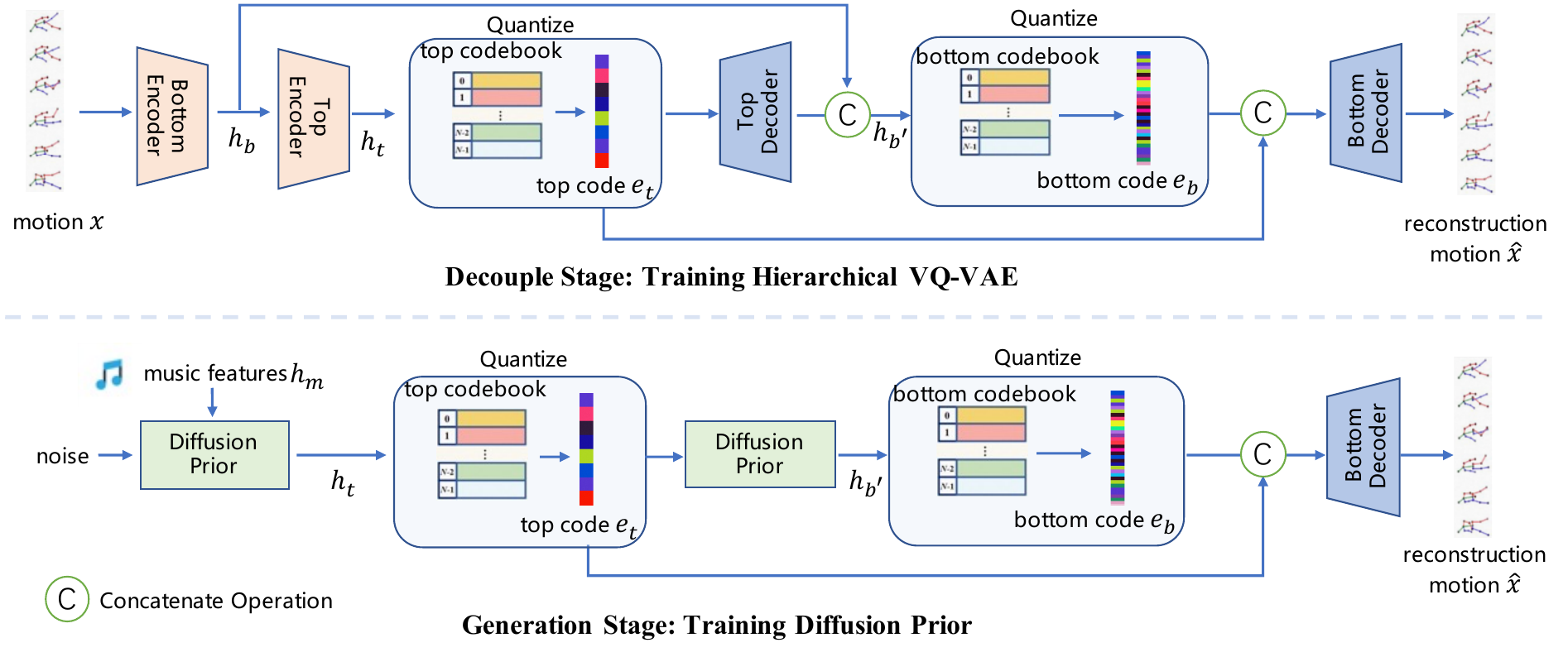}%\vspace{-0.2cm}
    \caption{Our method comprises two stages: the dance decouple stage and the dance generation stage. In the dance decouple stage, a hierarchical VQ-VAE is trained to decouple dance pose and dance movement by representing them with bottom code $\bm{e}_b$ and top code $\bm{e}_t$. In the motion generation phase, a diffusion model is employed as a prior to model the distribution of latent codes. The latent codes are then decoded into dance sequences by a motion decoder.}
    \label{fig:fig2_framework}%\vspace{-0.15cm}
\end{figure*}

\section{Related Works}
\label{sec:related_works}

%-----------------------------
\subsection{Music-to-Dance}

Music-to-dance is a sub-task within Human Motion Synthesis: the generation of human body movements that align with the characteristics of the music, such as its genre, rhythm, and intensity. 
% Here are some datasets related to Music2Dance: AIST++ \cite{li2021ai} is a 3D dance dataset which contains 3D motion reconstructed from real dancers paired with music. FineDance \cite{li2023finedance} contains music-dance paired data with fine-grained hand motions, fine-grained genres, and accurate posture. AIOZ-GDANCE \cite{le2023music} is a new large scale dataset for music-driven group dance generation.
A classic approach involves using Graph-based Motion Synthesis methods \cite{chen2021choreomaster, gao2022pc}, which segment the motions in the dataset into various nodes. Edges between nodes are added based on their similarity. When provided with input music, this task can be regarded as a retrieval task, where the corresponding motion nodes are selected by matching the speed and amplitude of character movements in motion segments to the rhythm and intensity of the music.
However, Graph-based methods have their limitations, this cutting-and-pasting approach lacks diversity. To address this issue, \cite{zhuang2022music2dance} uses auto-regressive generative model to generate 3D dance motions with high realism and diversity, \cite{li2021ai, pu2022music, li2022danceformer, sun2022you, yin2023dance} use transformer-based architecture to generate more diversified dance motion that are well-attuned to the input music. 
However, auto-regressive models suffer from the issue of error accumulation, making it prone to freezing when generating long sequences of dance motions. \cite{aristidou2022rhythm} try to generates long-term sequences of human motions by forming a global structure that respects a specific dance genre. \cite{huang2020dance} proposes a curriculum learning strategy to alleviate error accumulation of auto-regressive models in long motion sequence generation.

Human choreographers design dance motions from music by firstly devising multiple choreographic dance units and then arranging the dance unit sequence according to music's rhythm, melody and emotion \cite{blom1982intimate, minton2017choreography}. Inspired by this, \cite{lee2019dancing, ye2020choreonet} decompose a dance sequence into a series of basic dance units and then compose a dance by organizing multiple basic dancing movements seamlessly according to the input music. \cite{siyao2022bailando} utilizes VQ-VAE to encode and quantize 3D pose sequences and then employs a transformer based architecture to combine these discrete codes and generate corresponding motions.

\subsection{Motion Diffusion Models}

% Recently, the success of generative models in the text-to-image task has spurred the exploration of diffusion-based models in motion generation task. For instance, \cite{zhao2023taming, tseng2023edge} utilize diffusion models to generate more diverse dance movements. In the next section, more related works about motion diffusion models will be introduced.

Human motion generation, learned from motion capture data, can be conditioned by any signal that describes the motion, such as text and audio. 
For the text-to-motion task, \cite{tevet2022motionclip, hong2022avatarclip} represent motions as images and utilize the CLIP model \cite{radford2021learning} to align texts and images in latent space, enabling texts to control the generation of motions.
The success of generative models in the text-to-image task has spurred the exploration of diffusion-based models in motion generation task. \cite{zhang2022motiondiffuse, zhang2023remodiffuse, tevet2022human, chen2023executing, zhao2023modiff, kim2023flame} employ diffusion models to address the text-to-motion task, while \cite{zhao2023taming, tseng2023edge} use diffusion models to tackle the music-to-dance task.
Given that text, audio, and music all serve as conditions for motion generation task, \cite{zhou2023ude, dabral2023mofusion, ma2022pretrained, gong2023tm2d} address these sub-tasks simultaneously by designing a unified generative framework.
One of the challenges in motion generation is the problem of generating long sequences. \cite{tseng2023edge} generates choreographies of any length by imposing temporal constraints on batches of sequences. \cite{shafir2023human} generates arbitrarily-long sequences with text control per interval using a fixed motion diffusion prior. \cite{lee2023multiact} generates long-term 3D human motion from multiple action labels.

%-----------------------------
\subsection{Hierarchical VQ-VAE}
VQ-VAE \cite{van2017neural} is a generative model that leverages vector quantization to efficiently encode and decode high-dimensional data. The encoder learns a discrete latent representation, and the auto-regressive decoder is learnt as a prior to generate images.
To obtain higher-quality images, generative models often employ the coarse to fine strategy to generate images progressively. For example, StyleGAN \cite{karras2019style, karras2020analyzing} uses a synthesis network to progressively increase the feature resolutions and recovering high-resolution images from sampled noise. \cite{razavi2019generating, de2019hierarchical} introduce a hierarchical VQ-VAE structure comprising multiple nested encoders and decoders. In the field of music generation, \cite{dieleman2018challenge, dhariwal2020jukebox} also utilize the hierarchical VQ-VAE to represent more abstract musical features. A key feature of this kind of model is its use of multiple discrete codebooks during the encoding and decoding processes. The lower-level codebook captures fine-grained details, while the top-level codebook captures higher-level semantic information. 
% In this paper, we choose this hierarchical structure to represent specific dance poses and more abstract dance movements.

% TODO:这里需要补充VQ-VAE相关的论文和方法,可以参考Bailando等文章

% % 对于text2motion任务，[MotionCLIP, AvatarCLIP]把motion表示成图像，利用CLIP在隐空间将文本和图像对齐的方式，使得文本可以控制动作的生成。
% The success of generative models in the text-to-image task has spurred the exploration of diffusion-based models in motion generation task.
% [MotionDiffuse, ReMoDiffuse, MDM, Executing your Commands, Modiff, FLAME] 相继使用diffusion Model来解决text-to-motion的任务，[Taming, EDGE] 使用diffusion Model解决music-to-dance的任务，都取得了优异的结果。
% 由于文本和音频、音乐都是动作生成的条件，因此，[UDE, MoFusion, Pretrained diffusion Models]通过设计一个统一的生成框架，来同时解决这些子任务。
% 使用diffusion模型还需要解决的一个问题是生成长时的动作序列。[EDGE] generates choreographies of any length by imposing temporal constraints on batches of sequences. [PriorMDM] generates arbitrarily-long sequences with text control per interval using a fixed motion diffusion prior. [MultiAct] generates long-term 3D human motion from multiple action label

\newcommand{\h}{\bm{h}}
\newcommand{\x}{\bm{x}}
\newcommand{\E}{\bm{E}}
\newcommand{\D}{\bm{D}}
\newcommand{\N}{\bm{N}}
\newcommand{\e}{\bm{e}}
\newcommand{\w}{\bm{w}}

\section{Method}
\label{sec:method}
  
\subsection{Overview}
In the professional field of choreography, a dance phrase consists of several dance poses and dance movements. Dance poses composed of a series of basic meaningful body postures, while dance movements can reflect dynamic changes such as the rhythm, melody, and style of dance. 
Despite drawing inspiration from the knowledge mentioned above, it is still a challenge to model dance poses and dance movements. In real dance motions, poses and movements do not exhibit a straightforward one-to-one mapping, they are intricately intertwined and difficult to disentangle. 
We try to decouple and represent dance poses and dance movements explicitly by carefully selecting appropriate model structures and designing loss functions.

In Sec.\ref{section_3_3}, a hierarchical VQ-VAE architecture is used to disentangle and model dance poses and dance movements in different feature space levels, with the bottom latent code representing dance poses and the top latent code representing dance movements. The relevant experiments in Sec.\ref{latent_code_interp} demonstrated the interpretability of latent codes.
% With this disentanglement, we can not only represent dance motions more accurately, but also achieve applications such as dance style transfer and manual choreography.
In Sec.\ref{section_3_4}, we design a prior for modeling and sampling the latent codes of dance poses and dance movements by using a diffusion model. 
Music features are incorporated as conditions, ensuring that the generated dance motions match the music's type, rhythm, and intensity. The latent codes are fed into a decoder to reconstruct the dance motion sequence.

% In summary, we propose a two-stage dance generation framework called Hierarchical Latent Dance Diffuser (HLDD), which consists of two stages, i.e., phrase decouple stage and phrase generation stage. In phrase decouple stage, a hierarchical VQ-VAE is used to generate discrete motion latent codes known as top code and bottom code. In the phrase generation stage, a hierarchical VQ-Diffusion model is employed to model and sample these codes, and the final dance sequence is generated after passing through a Decoder.

\subsection{Data Representation}
Dance motions are represented by sequences of 3D human joints based on the SMPL model \cite{loper2023smpl}, which includes root's global positions and 6D rotation representation of 24 joints, resulting in a total dimension of $\x \in \mathbb{R}^{\N \times \w}$, where $\w \in \mathbb{R}^{3+24*6=147}$ is the dimension of motion features and $\N$ is the sequence number of frames.
The music features $\bm{\h_m}$ are extracted from the jukebox encoder \cite{dhariwal2020jukebox}, which is a large generative model trained on 1M songs and the features should have good representational ability for different types of music.
% The music features $\bm{m}$ comprises low-level features and high-level features. Librosa library \cite{mcfee2015librosa} is used to extract low-level features such as beats, onsets, and MFCCs. The jukebox encoder \cite{dhariwal2020jukebox} is used to extract high-level features.

\subsection{Hierarchical VQ-VAE}
\label{section_3_3}

The hierarchical VQ-VAE structure comprising multiple nested encoders and decoders. The lower-level features capture fine-grained details, while the top-level features capture higher-level semantic information.
For motion, its semantic manifestation occurs in the temporal domain. Short-term motion represents specific postures, while the continuous variation of motion over a longer duration can reflect the semantic information.
We use the hierarchical VQ-VAE and extract representations from intermediate layers to represent dance poses and dance movements.
During the encoding process, the dimension of features gradually increases to express more abstract characteristics. The temporal dimension compresses, enabling the representation of a broader temporal range.
Therefore, the low-level latent codes are used to represent dance poses, while the top-level latent codes are employed to represent dance movements.

Specifically, the motion input $\x$ passes through encoders $\E_{b}$ and  $\E_{t}$ to obtain bottom feature $\h_{b}$ and top feature $\h_{t}$, respectively. The top code $\e_{t}$ is obtained by the $Quantize$ operation \cite{van2017neural}, which are quantized vectors based on their distances to the prototype vectors in the codebook. 
Similarly, the bottom code $\e_{b}$ is the quantized results of $\h_b^{'}$, where $\h_b^{'}$ is the concatenate results of $\h_b$ and the result after passing through decoder $\D_t$ for $\e_t$. 
$\e_{b}$ and $\e_{t}$ are then concatenated and fed into the decoder $\D_b$ to get the reconstructed result $\hat{\x}$:
\begin{equation}
    \h_{b} = \E_{b}(\x), \h_{t} = \E_{t}(\h_b) \\
    \label{eq_1}
\end{equation}
\begin{equation}
    \e_{t} = Quantize(\h_{t}), \h_b^{'} = Concat(\D_t(\e_t), \h_{b})\\
    \label{eq_2}
\end{equation}
\begin{equation}
    \e_{b} = Quantize(\h_b^{'}), \hat{\x} = \D_b( Concat(\e_{t}, \e_{b}) )\\
    \label{eq_3}
\end{equation}
The whole pipeline is shown in Fig.\ref{fig:fig2_framework}.
For training the hierarchical VQ-VAE, the reconstruction loss with the commit losses are employeed:
\begin{equation}
    \begin{aligned}
        & \mathcal{L}_{VQ}(\x,\hat{\x}) = \Vert sg[\h_b^{'}] - \e_{b}\Vert _{2}^{2} + \alpha \Vert sg[\e_{b}] - \h_b^{'}\Vert _{2}^{2} \\
        & + \Vert sg[\h_{t}] - \e_{t}\Vert _{2}^{2} + \beta \Vert sg[\e_{t}] - \h_{t}\Vert _{2}^{2} + \Vert \x - \hat{\x}\Vert _{2}^{2}
    \end{aligned}
\end{equation}
% TODO：上面的公式需要增加2个超参数。考虑使用fi和cosai
In addition to using $L_2$ loss for motion reconstruction to train encoders, the corresponding codebooks for the two VQ-VAE networks are also trained simultaneously. And two commit losses are used to ensure that the embedding results of the encoders do not grow. Here the operator $sg$ refers to the stop-gradient operation that blocks gradients from flowing into its argument.

To improve physical realism in the absence of true simulation, we use four auxiliary losses: joint positions (Eq.\ref{loss_pos}), velocities (Eq.\ref{loss_vel}), accelerates (Eq.\ref{loss_acc}) and contact (Eq.\ref{loss_contact}) losses:

\begin{equation}
    \begin{aligned}
        \mathcal{L}_{pos} = \Vert  FK(\x) - FK(\hat{\x})  \Vert_{2}^{2}
    \end{aligned}
    \label{loss_pos}
\end{equation}
\begin{equation}
    \begin{aligned}
        \mathcal{L}_{vel} = \frac{1}{N-1}\sum_{i=1}^{N-1} \Vert(\x^{i+1}-\x^{i}) - (\hat{\x}^{i+1}-\hat{\x}^{i}) \Vert_{2}^{2}
    \end{aligned}
    \label{loss_vel}
\end{equation}
\begin{equation}
    \begin{aligned}
        \mathcal{L}_{acc} = \frac{1}{N-1}\sum_{i=1}^{N-1} \Vert(\bm{v}^{i+1}-\bm{v}^{i}) - (\hat{\bm{v}}^{i+1}-\hat{\bm{v}}^{i}) \Vert_{2}^{2}
    \end{aligned}
    \label{loss_acc}
\end{equation}
\begin{equation}
    \begin{aligned}
        \mathcal{L}_{contact} = \frac{1}{N-1}\sum_{i=1}^{N-1} \Vert ( FK(\hat{\x}^{i+1}) - FK(\hat{\x}^{i}) ) . \bm{b}^i \Vert_{2}^{2}
    \end{aligned}
    \label{loss_contact}
\end{equation}
where $FK(\cdot)$ denotes the forward kinematic function which converts joint angles into joint positions,$\bm{v}=\x^{i+1}-\x^{i}$, and $\bm{b}_i$ is the binary foot contact label of the pose at each frame $i$. The overall auxiliary loss is as follows:
\begin{equation}
    \begin{aligned}
        \mathcal{L}_{aux} = \mathcal{L}_{pos} + \gamma \mathcal{L}_{vel} + \phi \mathcal{L}_{acc} + \psi \mathcal{L}_{contact} 
    \end{aligned}
\end{equation}

In addition, we also use a modality alignment loss $\mathcal{L}_{MA}$ to align top code $\e_t$ which represents dance movements to high-level music features $\h_m$:
\begin{equation}
    \begin{aligned}
        \mathcal{L}_{MA} = \Vert \e_t - \h_m \Vert_{2}^{2}
    \end{aligned}
    \label{loss_MA}
\end{equation}
Final training loss is the weighted sum of the above losses:
\begin{equation}
    \begin{aligned}
        \mathcal{L} = \mathcal{L}_{VQ} + \lambda_{aux}\mathcal{L}_{aux} + \lambda_{MA}\mathcal{L}_{MA}
    \end{aligned}
\end{equation}

\subsection{Diffusion Model as Prior}
\label{section_3_4}
The next step is to learn a prior over the latent codes, thus motions can be generated from sampled latent codes conditioned with music features. 
Recently, the success of generative models in the text-to-image task has spurred the exploration of diffusion-based models in motion generation task. Diffusion models can learn the distribution of the latent space, aligning similar textual or music conditions with corresponding similar motion sequences. In addition to this, it has advantages in long-term dependency modeling and exhibits superior generation diversity.
Based on the above analysis of its advantages, we consider the transform based diffusion model as the prior. Two choices can be chosen, whether to use VQ-Diffusion \cite{gu2022vector} to model the distribution of the discrete latent codes $ p(\e_b, \e_t | \h_m)$, or use traditional gaussian diffusion to model the distribution of the continuous latent features $p(\h_b^{'}, \h_t |  \h_m)$ while preserving the codebooks. We've tried both and opt for the latter because it yields better results. 
% and the continuous gaussian process is easier to implement.

% \begin{figure}[htp]
%     \centering
%     \includegraphics[width=1\linewidth]{fig/figure_2.png}%\vspace{-0.2cm}
%     \caption{The Diffusion Model Prior Arthitecture.}
%     \label{fig:framework}%\vspace{-0.15cm}
% \end{figure}

\begin{figure}[htp]
    \centering
    \includegraphics[width=0.8\linewidth]{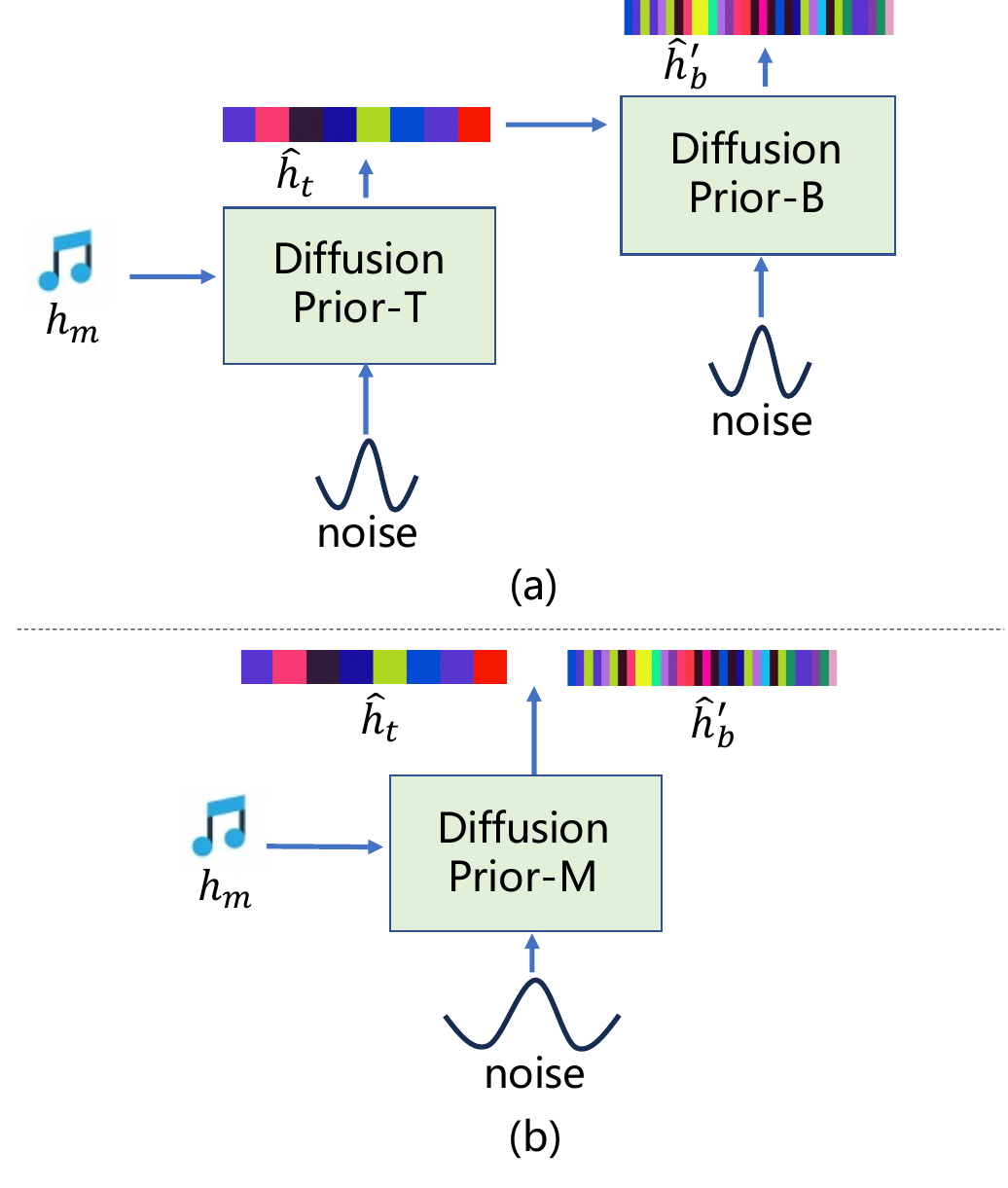}%\vspace{-0.2cm}
    \caption{Different prior model arthitecture. (a) Separatel model $p(\h_t | \h_m)$ and $p(\h_b^{'} | \h_t, \h_m)$. (b) Predict the joint probability $p(\h_b^{'}, \h_t |  \h_m)$ by concatenating $h_t$ and $h_b'$.}
    \label{fig:fig3_prior}%\vspace{-0.15cm}
\end{figure}

For the conditional joint probability distribution $p(\h_b^{'}, \h_t | \h_m)$, it can be further expressed as $p(\h_t | \h_m)p(\h_b^{'} | \h_t, \h_m)$. Therefore, we can use two diffusion models to separately model the conditional distribution $p(\h_t | \h_m)$ and $p(\h_b^{'} | \h_t, \h_m)$, as shown in Fig.\ref{fig:fig3_prior}(a). Another approach is to directly predict the joint probability distribution $p(\h_b^{'}, \h_t |  \h_m)$ by concatenating $h_t$ and $h_b'$, and then using a single diffusion model to model the joint probability distribution, as shown in Fig.\ref{fig:fig3_prior}(b). For detailed comparative experiments and results, please refer to Sec.\ref{ablation_study}.
Here we choose the latter for better clarification. We jointly model $\h_b^{'}$ and $\h_t$ by concatenating them: $\h = Concat(\h_b^{'}, \h_t)$, and follow the DDPM definition of diffusion as a Markov noising process, ${\{\h^{(t)}\}}^{T}_{t=0}$, where $\h^{(0)}$ is drawn from the data distributon and the forward process is defined as:
\begin{equation}
    \begin{aligned}
        q(\h^{(t)} | \h^{(t-1)}) \sim \mathcal{N} (\sqrt{1-\beta_t}\h^{(t-1)}, \beta_t\bm{I})
    \end{aligned} 
\end{equation}
where $\beta_t \in (0, 1)$ and $\bm{I}$ is the standard normal distribution. 
To approximate the distribution  $p(\h^{(0)} | \h_m)$, a transformer based neural network $\bm{G}$ is used to parameterize the chained reverse diffusion process $p_{\bm{G}}(\h^{(t-1)} | \h^{(t)}, \h_m )$ by directly predicting the clean sequence $\h^{(0)}$ as $\hat{\h}^{(0)} = \bm{G}(\h^{(t)}, t, \h_m)$. The generator $\bm{G}$ is optimized by the $L_2$ loss:
\begin{equation}
    \begin{aligned}
        \mathcal{L} = \mathbb{E}_{\h^{(0)}, t} [ \Vert \h^{(0)} - \hat{\h}^{(0)}) \Vert]_{2}^{2}
    \end{aligned} 
\end{equation}

In the inference stage, given the extracted music feature $\h_m$  as condition, we use DDIM to sample and gradually denoise $\h^{(t)}$ from noise, for timestep $t - 1$: $\hat{\h}^{(t-1)} \thicksim q(\bm{G}(\h^{(t)}, t, \h_m), t - 1)$. After predicting $\hat{\h}^{(0)}$, the final motion sequence can be obtained by the motion decoder.

\newcommand{\FIDk}{\textbf{$\rm{FID_k}\downarrow$}}  % Notation of FID_k score in Table.
\newcommand{\FIDg}{\textbf{$\rm{FID_g}\downarrow$}}  % Notation of FID_g score in Table.
\newcommand{\Divk}{\textbf{$\rm{Div_k}\rightarrow$}}  % Notation of Div_k score in Table.
\newcommand{\Divg}{\textbf{$\rm{Div_g}\rightarrow$}}  % Notation of Div_g score in Table.

\section{Experiments}
\label{sec:experiments}

\paragraph{Dataset.} 
We perform the training and evaluation on AIST++ dataset \cite{li2021ai}, which consists of 1,408 sequences of 3D human dance motion and are paired to music. The durations vary from 7.4 sec. to 48.0 sec. We re-use the train/test splits provided by the original dataset. 

% 跟原始数据集里面的测试集数量好像有些区别，原始数据集里面有40个，但是我用的好像只有10个
% resulting in total 40 unique choreographies in the test set. The train set is built by excluding all test musics and test choreographies from AIST++, resulting in total 329 unique choreographies in the train set.

\begin{table*}[t]
  \vspace{-0.2cm}
  \centering
  \begin{tabular}{lccccccc}
		\toprule
		\multirow{2}{*}{\textbf{Methods}} & \multicolumn{2}{c}{\textbf{Motion Quality}} & \multicolumn{2}{c}{\textbf{Motion Diversity}} & \multirow{2}{*}{\textbf{Beat Align.$\uparrow$}} & \multirow{2}{*}{\textbf{PFC$\downarrow$}} & \textbf{User Study} \\
		 \cmidrule(lr){2-3}\cmidrule(lr){4-5}\cmidrule(lr){8-8}& \FIDk & \FIDg & \Divk & \Divg & & & Ours WinRate\\
            % Methods & \FIDk & \FIDg & \Divk & \Divg & BAS & FPC \\
		\midrule
		%$512\times512$ resolution & & & &\\
            Ground Truth & 17.10 & 10.60 & 9.48 & 7.32 & 0.24 & 1.33 & 58.8\%\\
		FACT \cite{li2021ai} & 35.35 & 22.11 & 5.94 & 6.18 & 0.22 & 2.25 & 92.5\%\\
		Bailando \cite{siyao2022bailando} & 28.16 & 9.62 & 7.83 & 6.34 & 0.23 & 1.75 & 86.3\%\\
		EDGE \cite{tseng2023edge} & - & - & 10.68 & \textbf{7.62} & 0.27 & \textbf{1.65} & 64.6\%\\
		DanceMeld (Ours) & \textbf{22.74} & \textbf{9.18} & \textbf{8.74} & 7.89 & \textbf{0.28} & 1.72 & N/A\\
% 		\midrule
% 		$128\times128$ resolution & & & &\\
	
% 		\cmidrule(lr){1-1}ACGPN&115.7 &68.4&48.0 &39.2\\
% 		PF-AFN &86.6&29.4&49.5&\textbf{24.9}\\
% 		VITON-HD &95.0&27.4&\textbf{44.7}&25.6\\
% 		DGP (Ours) &\textbf{56.5}&\textbf{18.8}&46.8&29.9\\
		\bottomrule
	\end{tabular} \vspace{-0.25cm}
 \caption{
    Numerical metrics of FACT \cite{li2021ai}, Bailando \cite{siyao2022bailando}, EDGE \cite{tseng2023edge}, and our method on AIST++ dataset. $\downarrow$ indicates lower is better, $\rightarrow$ indicates closer to ground truth is better. Our method outperforms other approaches in terms of generated quality, as evidenced by the FID metric and validated by the User Study. In terms of diversity, our method remains on par with EDGE. Due to the incorporation of modality alignment loss and auxiliary losses, our method also exhibits advantages in terms of aligning with the beat of the music and PFC metrics.
  }
  \label{table:metrics}
\end{table*}

\paragraph{Implementation details.} 
We conduct pre-processing to segment the dance sequence to 512, the FPS is set at 60, the sliding window is 40. 
We use the features from the 36th layer of the Jukebox encoder, and a small encoder is employed to reduce the feature dimension to align with the top code. Same as motion input, the music feature's frame rate is also set at 60 FPS.
We employed a hierachical VQ-VAE with a total of 54M parameters. The downsample rate of the bottom encoder $\E_{b}$ and the top encoder $\E_{t}$ is 4 and 2, respectively. The dictionary size of the bottom codebook is set to 512, and 128 for the top codebook. The feature dimensions for both bottom code and top code is set to 512.
The hyperparameters $\alpha=\beta=0.02$, $\gamma=\phi=\psi=1$, $\lambda_{aux} = 1$, $\lambda_{MA} = 0.1$. Adam Optimizer \cite{kingma2014adam} is used with $lr=1e^{-4}$. The network is trained for 1000 epochs with a batch size of 64.
The diffusion process is implemented using DDPM with 1000 steps, by directly predicting $\h^{(0)}$ with a transformer encoder. The encoder has 8 layers, 8 attention heads, and a latent dimension of 512. For training the transformer encoder, a learning rate of $4e^{-4}$ is used with the Adan optimizer \cite{xie2022adan} over 500 epochs.
More details please refer to the supplementary materials.

\subsection{Comparison to Existing Methods} 

We compare our method against three state-of-the-art approaches: FACT \cite{li2021ai}, Bailando \cite{siyao2022bailando}, and EDGE \cite{tseng2023edge}. 
For evaluate metrics, we first extract kinetic feature and geometric feature, following the same steps as in \cite{li2021ai}. 
% The kinetic feature is defined on motion velocities and energies, which reflects the physical characteristics of dance, while the geometric feature is defined based on multiple manually templates of movements, which reflects the quality of choreography.
We evaluate the generated motion quality by calculating the distribution distance between the generated and the ground-truth motions using Fr\'echet Inception Distance (FID) on the extracted kinetic feature and geometric feature, $\rm{FID_k}$ and $\rm{FID_g}$, respectively.
For the motion diversity, we calculate the average Euclidean distance in the feature space across 40 generated motions on the AIST++ test set. The motion diversity in the geometric feature space and in the kinetic feature space are noted as $\rm{Div_k}$, $\rm{Div_g}$, respectively.
In addition, we also use Beat Alignment Score (BAS) \cite{li2021ai} to characterize the alignment between the dance motions and the musical rhythm. The Physical Foot Contact (PFC) score \cite{tseng2023edge} is also introduced to evaluate physical plausibility. 
Just as discussed in \cite{tseng2023edge}, due to the lack of a proficient motion encoder to represent the motion features, the FID metrics maybe unreliable as a measure of quality on this dataset. Thus human evaluations are introduced as a supplementary measure to qualitatively evaluate from a visual perspective. 
We hired 20 participants on an outsourcing platform for the User Study. We randomly selected 100 segments, each lasting 10 seconds, and had each participant make judgments. Each sample had results from our method and a randomly selected comparison method (including ground truth), both generated using the same music. Participants were asked to determine which dance result exhibited better generation quality and better alignment with the music.

The quantitative results are as shown in Tab. \ref{table:metrics}. 
Quantitatively, our method outperforms other approaches in terms of generated quality, as evidenced by the FID metric and validated by the User Study. In terms of diversity, our method remains on par with EDGE, which also relies on diffusion-based sampling. 
Due to the incorporation of modality alignment loss and auxiliary losses, our method also exhibits advantages in terms of aligning with the rhythm of the music and maintaining the authenticity of footstep movements.

\begin{table}[t]
  \vspace{-0.2cm}
  \centering
  \begin{tabular}{lcccc}
		\toprule
		\multirow{2}{*}{\textbf{Methods}} & \multicolumn{2}{c}{\textbf{Motion Quality}} & \multicolumn{2}{c}{\textbf{Motion Diversity}} \\
		 \cmidrule(lr){2-3}\cmidrule(lr){4-5}& \FIDk & \FIDg & \Divk & \Divg\\
		\midrule
		%$512\times512$ resolution & & & &\\
            Ours (Best) & \textbf{22.74} & \textbf{9.18} & \textbf{8.74} & \textbf{7.89}\\
            VQ-Diffusion & 35.35 & 22.11 & 5.94 & 6.18\\
            Single VQ-VAE & 25.48 & 11.75 & 7.96 & 6.12\\
            Prior-B+T & 28.16 & 9.62 & 7.83 & 6.34\\
% 		\midrule
% 		$128\times128$ resolution & & & &\\
	
% 		\cmidrule(lr){1-1}ACGPN&115.7 &68.4&48.0 &39.2\\
% 		PF-AFN &86.6&29.4&49.5&\textbf{24.9}\\
% 		VITON-HD &95.0&27.4&\textbf{44.7}&25.6\\
% 		DGP (Ours) &\textbf{56.5}&\textbf{18.8}&46.8&29.9\\
		\bottomrule
	\end{tabular} \vspace{-0.25cm}
 \caption{
    Ablation study on different network architectures. VQ-Diffusion predicts discrete latent codes distribution $ p(\e_b, \e_t | \h_m)$. Single VQ-VAE means not use hierarchical VQ-VAE. Prior-B+T refers to the architecture that use two diffusion models.} 
    % Prior-M refer to the architecture that using one single diffusion model to predict $p(\h_b^{'}, \h_t |  \h_m)$.}
  \label{table:ablation_prior}
\end{table}

\subsection{Ablation Studies} 
\label{ablation_study}
\paragraph{Network Selection.}

We conduct relevant ablation studies on the selection and design of network structures.
Our optimal model utilizes a hierarchical VQ-VAE architecture, employing the traditional gaussian diffusion to directly predict the joint probability distribution $p(\h_b^{'}, \h_t | \h_m)$.
When using VQ-Diffusion to predict the discrete probability distribution $ p(\e_b, \e_t | \h_m)$, we observed a significant decrease in the metrics. We suspect that the mask-and-replace diffusion strategy on VQ-Diffusion still has some limitations compared to gaussian diffusion process.
Next, we replace the hierarchical VQ-VAE model with a single VQ-VAE model and also observed some decrease in the metrics.
Our third comparative experiment involves using two diffusion models to predict conditional probability distributions for $p(\h_t | \h_m)$ and $p(\h_b^{'} | \h_t, \h_m)$, as shown in Fig.\ref{fig:fig3_prior}(b), the results were slightly inferior compared to directly predicting the joint probability distribution.
The corresponding results are shown in Tab. \ref{table:ablation_prior}.

\paragraph{Loss Terms.}
Here we discuss the impact of each loss term on the generated results. As seen in Tab.\ref{table:ablation_loss}, when auxiliary losses are not used, the PFC metric significantly deteriorates. Although $\rm{FID_k}$ shows a slight improvement, the generated motions exhibit minor fluctuations between time steps. These results suggest that contact loss may contribute to maintaining the authenticity of footstep movements, and the velocity and acceleration losses in the auxiliary losses contribute to temporal stability.

\begin{figure}[htp]
    \centering
    \includegraphics[width=0.9\linewidth]{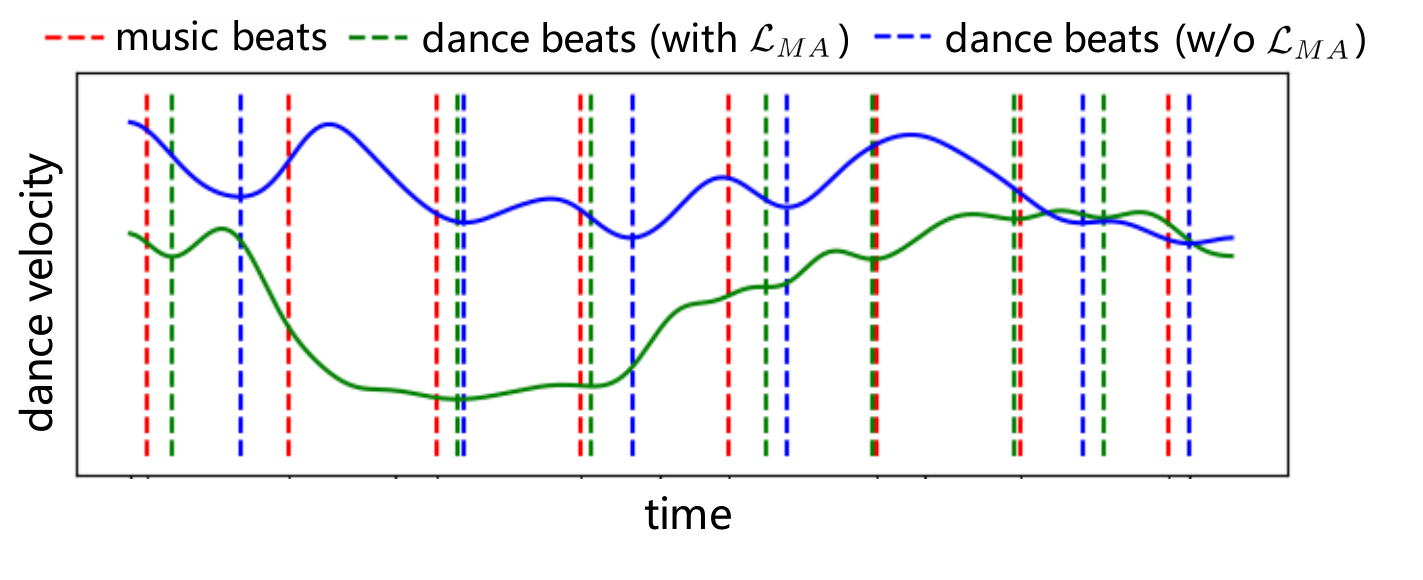}\vspace{-0.2cm}
    \caption{After using modality alignment loss, beats between dance motions and music are more synchronized.}
    \label{fig:fig4}\vspace{-0.15cm}
\end{figure}

The introduction of modality alignment loss enables the alignment of motion's top code and music features in the high-level latent space, allowing the top code to better represent music-related features, such as rhythm and melody. 
From  Tab.\ref{table:ablation_loss}, it can be observed that when modality alignment loss is not used, the BAS metric significantly decreases. Additionally, we randomly selected some in the wild music and extracted beat information from both music and motion. The visual results in Fig.\ref{fig:fig4} intuitively demonstrate that the modality alignment loss enhances the alignment of music and motion in terms of rhythm. 
% TODO:增加可视化的结果和

\begin{table}[t]
  \vspace{-0.2cm}
  \centering
\begin{tabular}{lcccc}
		\toprule
		\multirow{2}{*}{\textbf{Methods}} & \multicolumn{2}{c}{\textbf{Motion Quality}}  & \multirow{2}{*}{\textbf{Beat Align.$\uparrow$}} & \multirow{2}{*}{\textbf{PFC$\downarrow$}}  \\
		 \cmidrule(lr){2-3} & \FIDk & \FIDg  & & \\
            % Methods & \FIDk & \FIDg & \Divk & \Divg & BAS & FPC \\
		\midrule
		%$512\times512$ resolution & & & &\\
            Baseline & 22.74 & 9.18 & \textbf{0.28} & \textbf{1.72} \\
		w/o $\mathcal{L}_{aux}$ & \textbf{21.40} & 13.63  & 0.27 & 2.16 \\
		w/o $\mathcal{L}_{MA}$ & 27.90 & \textbf{8.91} & 0.23 & 1.84 \\
% 		\midrule
% 		$128\times128$ resolution & & & &\\
	
% 		\cmidrule(lr){1-1}ACGPN&115.7 &68.4&48.0 &39.2\\
% 		PF-AFN &86.6&29.4&49.5&\textbf{24.9}\\
% 		VITON-HD &95.0&27.4&\textbf{44.7}&25.6\\
% 		DGP (Ours) &\textbf{56.5}&\textbf{18.8}&46.8&29.9\\
		\bottomrule
	\end{tabular} \vspace{-0.25cm}
 \caption{
    The ablation study result of loss terms. The use of $\mathcal{L}_{aux}$ can improve physical realism  and enhance temporal stability. The use of $\mathcal{L}_{MA}$ improves the BAS metric.}
  \label{table:ablation_loss}
\end{table}

\subsection{The Interpretability of Latent Codes}
\label{latent_code_interp}
In our approach, bottom code can be explicitly represented as dance poses, while top code can characterize more abstract dance movements. We will validate the representational capabilities of the bottom code and top code through several visual experiments. 

\begin{figure}[htp]
    \centering
    \includegraphics[width=1.0\linewidth]{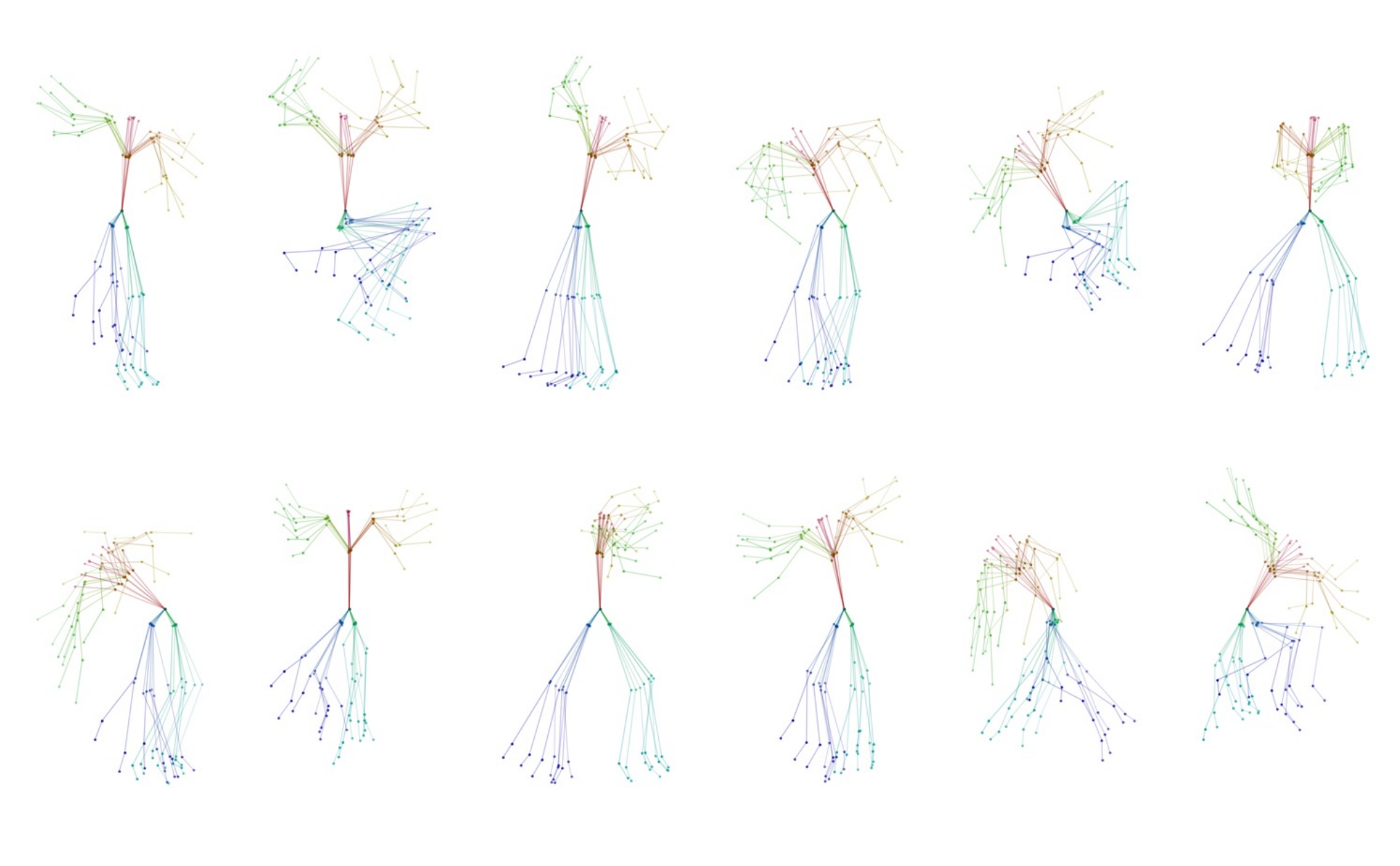}\vspace{-0.4cm}
    \caption{We present dance poses for 12 distinct bottom codes, combining each with a random selection of several top codes for visualization. The results reveal that the poses are constrained within a fixed range.}
    \label{fig:fig5}\vspace{-0.2cm}
\end{figure}

\paragraph{The Interpretability of Bottom Code.}
We design the following experiment to validate that bottom code can represent dance poses. By fixing the bottom code and changing the top code, we found that the generated motions were restrained within a fixed range. Fig.\ref{fig:fig5} shows 12 examples, each example represents the overlapped visual results of a fixed bottom code with different top codes. It is evident that the poses are restrained within a certain range. Different bottom codes correspond to different dance poses with clear distinctions. More visual results please refer to the supplementary materials.

\begin{figure}[htp]
    \centering
    \includegraphics[width=0.9\linewidth]{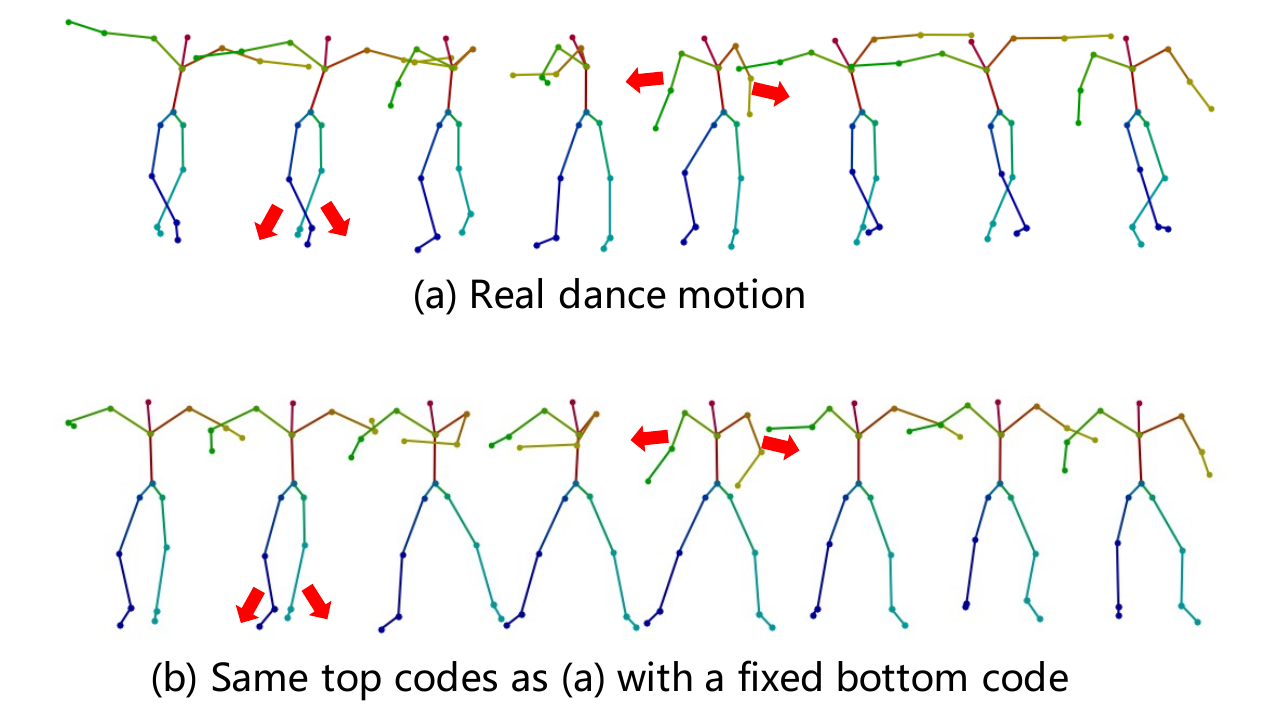}%\vspace{-0.2cm}
    \caption{(a) A real dance motion sequence. (b) Replace bottom code of (a) with a fixed value. Red arrows represent motion trends (dance movements). While the motion of (b) is constrained due to the fixed bottom code, it still maintains the same dance movements as (a), e.g., opening legs in the second column and lifting hands horizontally in the fifth column.}
    \label{fig:fig6}\vspace{-0.25cm}
\end{figure}

\begin{figure*}[htp]
    \centering
    \includegraphics[width=1.0\linewidth]{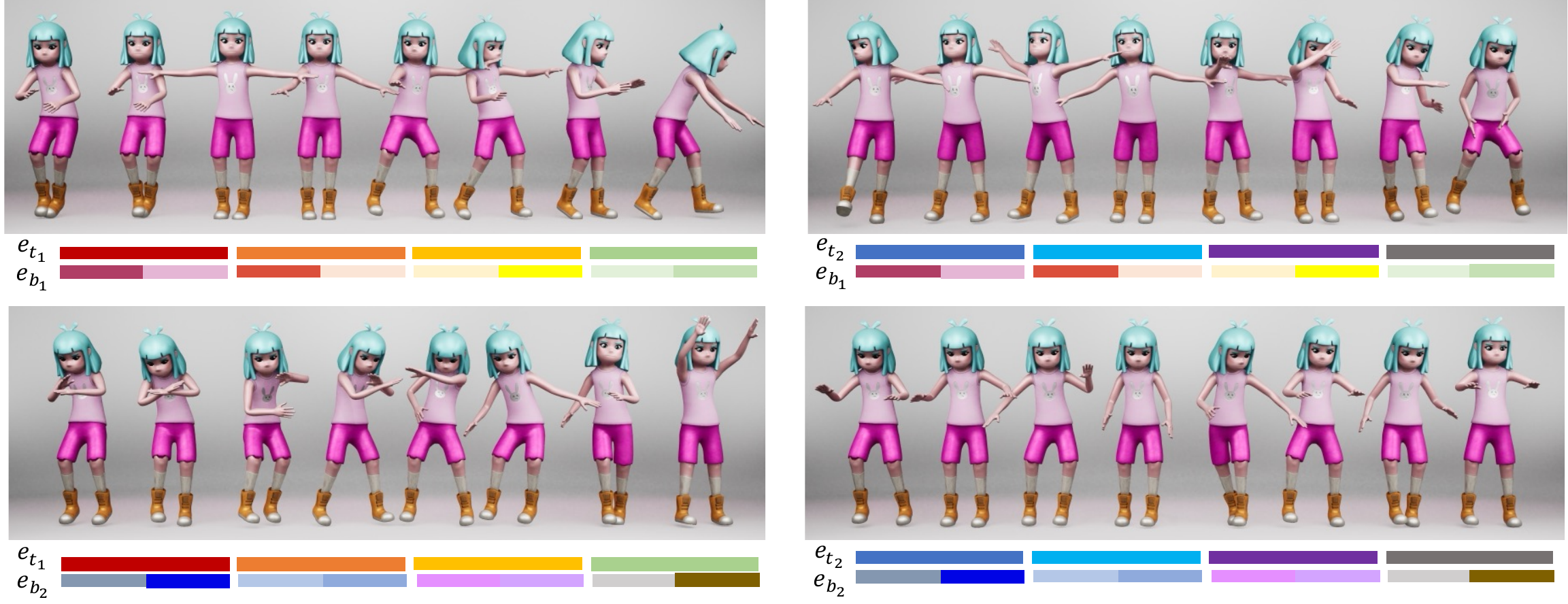}%\vspace{-0.2cm}
    \caption{Results of transferring bottom code and top code. $\e_{t_1}$ and  $\e_{t_2}$ represent two sets of top codes, $\e_{b_1}$ and $\e_{b_2}$ represent two sets of bottom codes. By modifying the top code (within the same row), the dance poses are retained, but the movements of the motion are altered. By modifying the bottom code (within the same column), the specific dance poses change, but the motion movements preserved.}
    \label{fig:fig7}%\vspace{-0.15cm}
\end{figure*}

\begin{figure}[htp]
    \centering
    \includegraphics[width=1.0\linewidth]{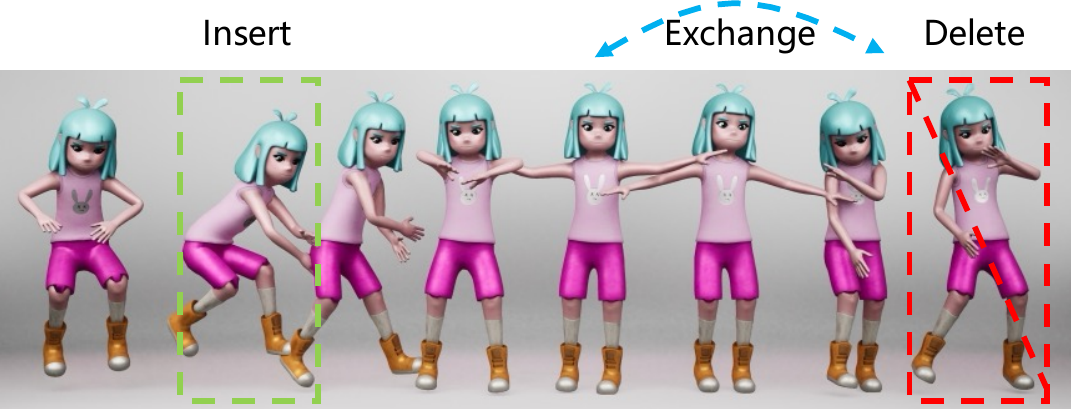}%\vspace{-0.2cm}
    \caption{Generated dance can be edited by inserting, deleting, replacing, or modifying the relative order of latent codes.}
    \label{fig:fig8}%\vspace{-0.15cm}
\end{figure}

\paragraph{The Interpretability of Top Code.}
As shown in Fig.\ref{fig:fig6}(a), we first select a segment of dance motion sequence and display it in the time dimension. Next, we replace the bottom code of this dance segment with a fixed value, as illustrated in Fig.\ref{fig:fig6}(b). Since the bottom code is replaced with a fixed value, the motion in Fig.\ref{fig:fig6}(b) is constrained within a certain range. However, from the temporal perspective, the motion trends (dance movements) of the two are consistent, e.g., opening legs in the second column and lifting hands horizontally in the fifth column (red arrows represent motion trends), which demonstrates that the top code can characterize dance movements. Please refer to the supplementary materials for more visual results.

\subsection{Applications}
Through the above verification of the interpretability of top code and bottom code, we've confirmed that the bottom code has the ability to represent dance poses, while the top code can signify dance movements. Using a hierarchical VQ-VAE structure, we have successfully decoupled dance pose and dance movement, enabling diverse applications. For example, by transferring different top codes, we can control the style, speed, and motion trends of the dance, as illustrated in Fig.\ref{fig:fig7}, $\e_{t_1}$ and  $\e_{t_2}$ represent two sets of top codes, $\e_{b_1}$ and $\e_{b_2}$ represent two sets of bottom codes. Different colors indicate different latent codes.By modifying the top code (within the same row), the dance poses are retained, but the speed and movements of the motion are altered. By modifying the bottom code (within the same column), the specific dance poses change, but the motion trends and dance movements are preserved. Furthermore, we can use this capability to achieve choreography editing. As illustrated in Fig.\ref{fig:fig8}, since the latent code corresponds to a dance phrase in a specific time segment, we can edit the generated dance by inserting, deleting, replacing, or modifying the relative order of latent codes.

\section{Limitations and Future Works}
\label{sec:limitation_and_future_works}
Generating long-term and temporally coherent dance sequences remains a challenge. While our method can generate dance sequences of any length by applying temporal constraints to batches \cite{tseng2023edge}, there is still a need to explore a way to model the global consistency like \cite{aristidou2022rhythm}, reflecting the concept of dance sections. In addition, the efficiency of diffusion-based methods is relatively low, preventing real-time generation. Consistency models\cite{song2023consistency, luo2023latent} may be a potential improvement to accelerate the generation process.

% Generating long and temporally coherent dance sequences poses a challenge. While our method can create dance sequences of any length by applying temporal constraints to batches, there is a need to explore modeling global consistency, akin to A, which captures the idea of dance sections. Moreover, diffusion-based methods currently have limited efficiency, hindering real-time generation. Consistency models could be a potential enhancement to expedite the generation process.
\section{Conclusions}
\label{sec:conclusion}
Our method draws inspiration from the domain of choreography and introduces a two-stage generative framework named DanceMeld. In the decouple stage, a hierarchical VQ-VAE is trained to obtain the bottom code and top code, representing dance pose and dance movement. In the generation stage, a diffusion model serves as a prior for modeling and sampling latent codes. 
We conduct several experiments to demonstrate the interpretability of bottom code and top code, demonstrating applications such as dance style transfer and dance editing. Validation on the AIST++ dataset reveals state-of-the-art performance, supported by both qualitative and quantitative results.

{
    \small
    \bibliographystyle{ieeenat_fullname}
    \bibliography{main}

\begin{thebibliography}{48}
\providecommand{\natexlab}[1]{#1}
\providecommand{\url}[1]{\texttt{#1}}
\expandafter\ifx\csname urlstyle\endcsname\relax
  \providecommand{\doi}[1]{doi: #1}\else
  \providecommand{\doi}{doi: \begingroup \urlstyle{rm}\Url}\fi

\bibitem[Alexanderson et~al.(2023)Alexanderson, Nagy, Beskow, and Henter]{alexanderson2023listen}
Simon Alexanderson, Rajmund Nagy, Jonas Beskow, and Gustav~Eje Henter.
\newblock Listen, denoise, action! audio-driven motion synthesis with diffusion models.
\newblock \emph{ACM Transactions on Graphics (TOG)}, 42\penalty0 (4):\penalty0 1--20, 2023.

\bibitem[Aristidou et~al.(2022)Aristidou, Yiannakidis, Aberman, Cohen-Or, Shamir, and Chrysanthou]{aristidou2022rhythm}
Andreas Aristidou, Anastasios Yiannakidis, Kfir Aberman, Daniel Cohen-Or, Ariel Shamir, and Yiorgos Chrysanthou.
\newblock Rhythm is a dancer: Music-driven motion synthesis with global structure.
\newblock \emph{IEEE Transactions on Visualization and Computer Graphics}, 2022.

\bibitem[Blom and Chaplin(1982)]{blom1982intimate}
Lynne~Anne Blom and L~Tarin Chaplin.
\newblock \emph{The intimate act of choreography}.
\newblock University of Pittsburgh Pre, 1982.

\bibitem[Chen et~al.(2021)Chen, Tan, Lei, Zhang, Guo, Zhang, and Hu]{chen2021choreomaster}
Kang Chen, Zhipeng Tan, Jin Lei, Song-Hai Zhang, Yuan-Chen Guo, Weidong Zhang, and Shi-Min Hu.
\newblock Choreomaster: choreography-oriented music-driven dance synthesis.
\newblock \emph{ACM Transactions on Graphics (TOG)}, 40\penalty0 (4):\penalty0 1--13, 2021.

\bibitem[Chen et~al.(2023)Chen, Jiang, Liu, Huang, Fu, Chen, and Yu]{chen2023executing}
Xin Chen, Biao Jiang, Wen Liu, Zilong Huang, Bin Fu, Tao Chen, and Gang Yu.
\newblock Executing your commands via motion diffusion in latent space.
\newblock In \emph{Proceedings of the IEEE/CVF Conference on Computer Vision and Pattern Recognition}, pages 18000--18010, 2023.

\bibitem[Dabral et~al.(2023)Dabral, Mughal, Golyanik, and Theobalt]{dabral2023mofusion}
Rishabh Dabral, Muhammad~Hamza Mughal, Vladislav Golyanik, and Christian Theobalt.
\newblock Mofusion: A framework for denoising-diffusion-based motion synthesis.
\newblock In \emph{Proceedings of the IEEE/CVF Conference on Computer Vision and Pattern Recognition}, pages 9760--9770, 2023.

\bibitem[De~Fauw et~al.(2019)De~Fauw, Dieleman, and Simonyan]{de2019hierarchical}
Jeffrey De~Fauw, Sander Dieleman, and Karen Simonyan.
\newblock Hierarchical autoregressive image models with auxiliary decoders.
\newblock \emph{arXiv preprint arXiv:1903.04933}, 2019.

\bibitem[Dhariwal et~al.(2020)Dhariwal, Jun, Payne, Kim, Radford, and Sutskever]{dhariwal2020jukebox}
Prafulla Dhariwal, Heewoo Jun, Christine Payne, Jong~Wook Kim, Alec Radford, and Ilya Sutskever.
\newblock Jukebox: A generative model for music.
\newblock \emph{arXiv preprint arXiv:2005.00341}, 2020.

\bibitem[Dieleman et~al.(2018)Dieleman, van~den Oord, and Simonyan]{dieleman2018challenge}
Sander Dieleman, Aaron van~den Oord, and Karen Simonyan.
\newblock The challenge of realistic music generation: modelling raw audio at scale.
\newblock \emph{Advances in neural information processing systems}, 31, 2018.

\bibitem[Gao et~al.(2022)Gao, Pu, Zhang, Shan, and Zheng]{gao2022pc}
Jibin Gao, Junfu Pu, Honglun Zhang, Ying Shan, and Wei-Shi Zheng.
\newblock Pc-dance: Posture-controllable music-driven dance synthesis.
\newblock In \emph{Proceedings of the 30th ACM International Conference on Multimedia}, pages 1261--1269, 2022.

\bibitem[Gong et~al.(2023)Gong, Lian, Chang, Guo, Jiang, Zuo, Mi, and Wang]{gong2023tm2d}
Kehong Gong, Dongze Lian, Heng Chang, Chuan Guo, Zihang Jiang, Xinxin Zuo, Michael~Bi Mi, and Xinchao Wang.
\newblock Tm2d: Bimodality driven 3d dance generation via music-text integration.
\newblock In \emph{Proceedings of the IEEE/CVF International Conference on Computer Vision}, pages 9942--9952, 2023.

\bibitem[Gu et~al.(2022)Gu, Chen, Bao, Wen, Zhang, Chen, Yuan, and Guo]{gu2022vector}
Shuyang Gu, Dong Chen, Jianmin Bao, Fang Wen, Bo Zhang, Dongdong Chen, Lu Yuan, and Baining Guo.
\newblock Vector quantized diffusion model for text-to-image synthesis.
\newblock In \emph{Proceedings of the IEEE/CVF Conference on Computer Vision and Pattern Recognition}, pages 10696--10706, 2022.

\bibitem[Hong et~al.(2022)Hong, Zhang, Pan, Cai, Yang, and Liu]{hong2022avatarclip}
Fangzhou Hong, Mingyuan Zhang, Liang Pan, Zhongang Cai, Lei Yang, and Ziwei Liu.
\newblock Avatarclip: Zero-shot text-driven generation and animation of 3d avatars.
\newblock \emph{arXiv preprint arXiv:2205.08535}, 2022.

\bibitem[Huang et~al.(2020)Huang, Hu, Wu, Sawada, Zhang, and Jiang]{huang2020dance}
Ruozi Huang, Huang Hu, Wei Wu, Kei Sawada, Mi Zhang, and Daxin Jiang.
\newblock Dance revolution: Long-term dance generation with music via curriculum learning.
\newblock \emph{arXiv preprint arXiv:2006.06119}, 2020.

\bibitem[Karras et~al.(2019)Karras, Laine, and Aila]{karras2019style}
Tero Karras, Samuli Laine, and Timo Aila.
\newblock A style-based generator architecture for generative adversarial networks.
\newblock In \emph{Proceedings of the IEEE/CVF conference on computer vision and pattern recognition}, pages 4401--4410, 2019.

\bibitem[Karras et~al.(2020)Karras, Laine, Aittala, Hellsten, Lehtinen, and Aila]{karras2020analyzing}
Tero Karras, Samuli Laine, Miika Aittala, Janne Hellsten, Jaakko Lehtinen, and Timo Aila.
\newblock Analyzing and improving the image quality of stylegan.
\newblock In \emph{Proceedings of the IEEE/CVF conference on computer vision and pattern recognition}, pages 8110--8119, 2020.

\bibitem[Kim et~al.(2022)Kim, Oh, Kim, Tong, and Lee]{kim2022brand}
Jinwoo Kim, Heeseok Oh, Seongjean Kim, Hoseok Tong, and Sanghoon Lee.
\newblock A brand new dance partner: Music-conditioned pluralistic dancing controlled by multiple dance genres.
\newblock In \emph{Proceedings of the IEEE/CVF Conference on Computer Vision and Pattern Recognition}, pages 3490--3500, 2022.

\bibitem[Kim et~al.(2023)Kim, Kim, and Choi]{kim2023flame}
Jihoon Kim, Jiseob Kim, and Sungjoon Choi.
\newblock Flame: Free-form language-based motion synthesis \& editing.
\newblock In \emph{Proceedings of the AAAI Conference on Artificial Intelligence}, pages 8255--8263, 2023.

\bibitem[Kingma and Ba(2014)]{kingma2014adam}
Diederik~P Kingma and Jimmy Ba.
\newblock Adam: A method for stochastic optimization.
\newblock \emph{arXiv preprint arXiv:1412.6980}, 2014.

\bibitem[Lee et~al.(2019)Lee, Yang, Liu, Wang, Lu, Yang, and Kautz]{lee2019dancing}
Hsin-Ying Lee, Xiaodong Yang, Ming-Yu Liu, Ting-Chun Wang, Yu-Ding Lu, Ming-Hsuan Yang, and Jan Kautz.
\newblock Dancing to music.
\newblock \emph{Advances in neural information processing systems}, 32, 2019.

\bibitem[Lee et~al.(2023)Lee, Moon, and Lee]{lee2023multiact}
Taeryung Lee, Gyeongsik Moon, and Kyoung~Mu Lee.
\newblock Multiact: Long-term 3d human motion generation from multiple action labels.
\newblock In \emph{Proceedings of the AAAI Conference on Artificial Intelligence}, pages 1231--1239, 2023.

\bibitem[Li et~al.(2022)Li, Zhao, Zhelun, and Sheng]{li2022danceformer}
Buyu Li, Yongchi Zhao, Shi Zhelun, and Lu Sheng.
\newblock Danceformer: Music conditioned 3d dance generation with parametric motion transformer.
\newblock In \emph{Proceedings of the AAAI Conference on Artificial Intelligence}, pages 1272--1279, 2022.

\bibitem[Li et~al.(2021)Li, Yang, Ross, and Kanazawa]{li2021ai}
Ruilong Li, Shan Yang, David~A Ross, and Angjoo Kanazawa.
\newblock Ai choreographer: Music conditioned 3d dance generation with aist++.
\newblock In \emph{Proceedings of the IEEE/CVF International Conference on Computer Vision}, pages 13401--13412, 2021.

\bibitem[Li et~al.(2023)Li, Zhao, Zhang, Su, Ren, Zhang, Tang, and Li]{li2023finedance}
Ronghui Li, Junfan Zhao, Yachao Zhang, Mingyang Su, Zeping Ren, Han Zhang, Yansong Tang, and Xiu Li.
\newblock Finedance: A fine-grained choreography dataset for 3d full body dance generation.
\newblock In \emph{Proceedings of the IEEE/CVF International Conference on Computer Vision}, pages 10234--10243, 2023.

\bibitem[Loper et~al.(2023)Loper, Mahmood, Romero, Pons-Moll, and Black]{loper2023smpl}
Matthew Loper, Naureen Mahmood, Javier Romero, Gerard Pons-Moll, and Michael~J Black.
\newblock Smpl: A skinned multi-person linear model.
\newblock In \emph{Seminal Graphics Papers: Pushing the Boundaries, Volume 2}, pages 851--866. 2023.

\bibitem[Luo et~al.(2023)Luo, Tan, Huang, Li, and Zhao]{luo2023latent}
Simian Luo, Yiqin Tan, Longbo Huang, Jian Li, and Hang Zhao.
\newblock Latent consistency models: Synthesizing high-resolution images with few-step inference, 2023.

\bibitem[Ma et~al.(2022)Ma, Bai, and Zhou]{ma2022pretrained}
Jianxin Ma, Shuai Bai, and Chang Zhou.
\newblock Pretrained diffusion models for unified human motion synthesis.
\newblock \emph{arXiv preprint arXiv:2212.02837}, 2022.

\bibitem[Minton(2017)]{minton2017choreography}
Sandra~Cerny Minton.
\newblock \emph{Choreography: a basic approach using improvisation}.
\newblock Human Kinetics, 2017.

\bibitem[Pu and Shan(2022)]{pu2022music}
Junfu Pu and Ying Shan.
\newblock Music-driven dance regeneration with controllable key pose constraints.
\newblock \emph{arXiv preprint arXiv:2207.03682}, 2022.

\bibitem[Radford et~al.(2021)Radford, Kim, Hallacy, Ramesh, Goh, Agarwal, Sastry, Askell, Mishkin, Clark, et~al.]{radford2021learning}
Alec Radford, Jong~Wook Kim, Chris Hallacy, Aditya Ramesh, Gabriel Goh, Sandhini Agarwal, Girish Sastry, Amanda Askell, Pamela Mishkin, Jack Clark, et~al.
\newblock Learning transferable visual models from natural language supervision.
\newblock In \emph{International conference on machine learning}, pages 8748--8763. PMLR, 2021.

\bibitem[Razavi et~al.(2019)Razavi, Van~den Oord, and Vinyals]{razavi2019generating}
Ali Razavi, Aaron Van~den Oord, and Oriol Vinyals.
\newblock Generating diverse high-fidelity images with vq-vae-2.
\newblock \emph{Advances in neural information processing systems}, 32, 2019.

\bibitem[Shafir et~al.(2023)Shafir, Tevet, Kapon, and Bermano]{shafir2023human}
Yonatan Shafir, Guy Tevet, Roy Kapon, and Amit~H Bermano.
\newblock Human motion diffusion as a generative prior.
\newblock \emph{arXiv preprint arXiv:2303.01418}, 2023.

\bibitem[Siyao et~al.(2022)Siyao, Yu, Gu, Lin, Wang, Qian, Loy, and Liu]{siyao2022bailando}
Li Siyao, Weijiang Yu, Tianpei Gu, Chunze Lin, Quan Wang, Chen Qian, Chen~Change Loy, and Ziwei Liu.
\newblock Bailando: 3d dance generation by actor-critic gpt with choreographic memory.
\newblock In \emph{Proceedings of the IEEE/CVF Conference on Computer Vision and Pattern Recognition}, pages 11050--11059, 2022.

\bibitem[Song et~al.(2023)Song, Dhariwal, Chen, and Sutskever]{song2023consistency}
Yang Song, Prafulla Dhariwal, Mark Chen, and Ilya Sutskever.
\newblock Consistency models.
\newblock 2023.

\bibitem[Sun et~al.(2022)Sun, Wang, Hu, Lai, Jin, and Hu]{sun2022you}
Jiangxin Sun, Chunyu Wang, Huang Hu, Hanjiang Lai, Zhi Jin, and Jian-Fang Hu.
\newblock You never stop dancing: Non-freezing dance generation via bank-constrained manifold projection.
\newblock \emph{Advances in Neural Information Processing Systems}, 35:\penalty0 9995--10007, 2022.

\bibitem[Tevet et~al.(2022{\natexlab{a}})Tevet, Gordon, Hertz, Bermano, and Cohen-Or]{tevet2022motionclip}
Guy Tevet, Brian Gordon, Amir Hertz, Amit~H Bermano, and Daniel Cohen-Or.
\newblock Motionclip: Exposing human motion generation to clip space.
\newblock In \emph{European Conference on Computer Vision}, pages 358--374. Springer, 2022{\natexlab{a}}.

\bibitem[Tevet et~al.(2022{\natexlab{b}})Tevet, Raab, Gordon, Shafir, Cohen-Or, and Bermano]{tevet2022human}
Guy Tevet, Sigal Raab, Brian Gordon, Yonatan Shafir, Daniel Cohen-Or, and Amit~H Bermano.
\newblock Human motion diffusion model.
\newblock \emph{arXiv preprint arXiv:2209.14916}, 2022{\natexlab{b}}.

\bibitem[Tseng et~al.(2023)Tseng, Castellon, and Liu]{tseng2023edge}
Jonathan Tseng, Rodrigo Castellon, and Karen Liu.
\newblock Edge: Editable dance generation from music.
\newblock In \emph{Proceedings of the IEEE/CVF Conference on Computer Vision and Pattern Recognition}, pages 448--458, 2023.

\bibitem[Van Den~Oord et~al.(2017)Van Den~Oord, Vinyals, et~al.]{van2017neural}
Aaron Van Den~Oord, Oriol Vinyals, et~al.
\newblock Neural discrete representation learning.
\newblock \emph{Advances in neural information processing systems}, 30, 2017.

\bibitem[Xie et~al.(2022)Xie, Zhou, Li, Lin, and Yan]{xie2022adan}
Xingyu Xie, Pan Zhou, Huan Li, Zhouchen Lin, and Shuicheng Yan.
\newblock Adan: Adaptive nesterov momentum algorithm for faster optimizing deep models.
\newblock \emph{arXiv preprint arXiv:2208.06677}, 2022.

\bibitem[Ye et~al.(2020)Ye, Wu, Jia, Bu, Chen, Meng, and Wang]{ye2020choreonet}
Zijie Ye, Haozhe Wu, Jia Jia, Yaohua Bu, Wei Chen, Fanbo Meng, and Yanfeng Wang.
\newblock Choreonet: Towards music to dance synthesis with choreographic action unit.
\newblock In \emph{Proceedings of the 28th ACM International Conference on Multimedia}, pages 744--752, 2020.

\bibitem[Yin et~al.(2023)Yin, Yin, Baraka, Kragic, and Bj{\"o}rkman]{yin2023dance}
Wenjie Yin, Hang Yin, Kim Baraka, Danica Kragic, and M{\aa}rten Bj{\"o}rkman.
\newblock Dance style transfer with cross-modal transformer.
\newblock In \emph{Proceedings of the IEEE/CVF Winter Conference on Applications of Computer Vision}, pages 5058--5067, 2023.

\bibitem[Zhang et~al.(2022)Zhang, Cai, Pan, Hong, Guo, Yang, and Liu]{zhang2022motiondiffuse}
Mingyuan Zhang, Zhongang Cai, Liang Pan, Fangzhou Hong, Xinying Guo, Lei Yang, and Ziwei Liu.
\newblock Motiondiffuse: Text-driven human motion generation with diffusion model.
\newblock \emph{arXiv preprint arXiv:2208.15001}, 2022.

\bibitem[Zhang et~al.(2023)Zhang, Guo, Pan, Cai, Hong, Li, Yang, and Liu]{zhang2023remodiffuse}
Mingyuan Zhang, Xinying Guo, Liang Pan, Zhongang Cai, Fangzhou Hong, Huirong Li, Lei Yang, and Ziwei Liu.
\newblock Remodiffuse: Retrieval-augmented motion diffusion model.
\newblock \emph{arXiv preprint arXiv:2304.01116}, 2023.

\bibitem[Zhao et~al.(2023{\natexlab{a}})Zhao, Liu, Ren, Dai, and Sebe]{zhao2023modiff}
Mengyi Zhao, Mengyuan Liu, Bin Ren, Shuling Dai, and Nicu Sebe.
\newblock Modiff: Action-conditioned 3d motion generation with denoising diffusion probabilistic models.
\newblock \emph{arXiv preprint arXiv:2301.03949}, 2023{\natexlab{a}}.

\bibitem[Zhao et~al.(2023{\natexlab{b}})Zhao, Bai, Chen, Wang, and Pan]{zhao2023taming}
Zhuoran Zhao, Jinbin Bai, Delong Chen, Debang Wang, and Yubo Pan.
\newblock Taming diffusion models for music-driven conducting motion generation.
\newblock \emph{arXiv preprint arXiv:2306.10065}, 2023{\natexlab{b}}.

\bibitem[Zhou and Wang(2023)]{zhou2023ude}
Zixiang Zhou and Baoyuan Wang.
\newblock Ude: A unified driving engine for human motion generation.
\newblock In \emph{Proceedings of the IEEE/CVF Conference on Computer Vision and Pattern Recognition}, pages 5632--5641, 2023.

\bibitem[Zhuang et~al.(2022)Zhuang, Wang, Chai, Wang, Shao, and Xia]{zhuang2022music2dance}
Wenlin Zhuang, Congyi Wang, Jinxiang Chai, Yangang Wang, Ming Shao, and Siyu Xia.
\newblock Music2dance: Dancenet for music-driven dance generation.
\newblock \emph{ACM Transactions on Multimedia Computing, Communications, and Applications (TOMM)}, 18\penalty0 (2):\penalty0 1--21, 2022.

\end{thebibliography}
}

% WARNING: do not forget to delete the supplementary pages from your submission 
% \input{sec/X_suppl}

\end{document}

%%%%%%%%% TITLE - PLEASE UPDATE
\title{DanceMeld: Unraveling Dance Phrases with Hierarchical Latent Codes for Music-to-Dance Synthesis}  % **** Enter the paper title here

\clearpage
\setcounter{page}{1}
\maketitlesupplementary

\section{Algorithm}
\label{sec:algorithm}
% 
% Our method DanceMeld comprises two stages: the dance decouple stage and the dance generation stage.
% We use pseudocode to provide a summarized description of the steps in each stage.
We use pseudocode to provide a summarized description of the decouple and the generation stage.
\begin{algorithm}[h]
	\caption{Decouple stage (training hierarchical VQ-VAE)}
	\label{algorithm}
	\begin{algorithmic}
		\STATE {\bfseries Require:} Functions $\E_b$, $\E_t$, $\D_b$, $\D_t$, $\x$, $\h_m$. 
            %\vspace{-0.2cm}
		\REPEAT
            
            \STATE $\h_b \gets \E_b(\x), \h_t \gets \E_t(\h_b)$ \\
            \COMMENT{get encoder features}\\
            \STATE $\e_{t} \gets Quantize(\h_{t})$ \\
            \COMMENT{get top code}\\
            \STATE $\h_b^{'} \gets Concat(\D_t(\e_t), \h_{b})$\\
            \STATE $\e_{b} \gets Quantize(\h_b^{'})$ \\
            \COMMENT{get bottom code}\\
            \STATE $\hat{\x} = \D_b( Concat(\e_{t}, \e_{b}) )$\\
            \STATE $\theta = Update(\mathcal{L}(\x, \hat{\x}))$\\

		% \STATE Compute the gradient of $f$ at $\bm{w}_k$ as $\nabla f(\bm{w}_k)$.
		% \STATE Update $\bm{w}_k$ by%\vspace{-0.3cm}
		% \begin{equation}
		%     \bm{w}_{k+1}=\bm{w}_k-\gamma\nabla f(\bm{w}_k).%\vspace{-0.5cm}
		% \end{equation}
		% \STATE Project $\bm{w}_{k+1}$ back to $\mathcal{C}$ by %\vspace{-0.3cm}
		% \begin{equation}\label{eq:projection}
		%     \bm{w}_{k+1}=\arg\min_{\bm{w}\in\mathcal{C}}\Vert \bm{w}_{k+1}-\bm{w}\Vert.%\vspace{-0.5cm}
		% \end{equation}
		% \STATE Update counter as $k = k+1$.
		\UNTIL{Convergence.}
	\end{algorithmic}
\end{algorithm}
\begin{algorithm}[h]
	\caption{Generation stage (training diffusion prior)}
	\label{algorithm}
	\begin{algorithmic}
		\STATE $H, H_m \gets \emptyset$\\
            \COMMENT{training set}\\

            \FOR {($\x$, $\h_m$) in training\_set}
                \STATE Get $\h_t, \h_b^{'}$ according to Algorithm 1
                \STATE $\h \gets Concat(\h_t, \h_b^{'})$
                \STATE $H \gets H \cup \h$
                \STATE $H_m \gets H_m \cup \h_m$
            \ENDFOR
            \STATE $p(\h_b^{'}, \h_t | \h_m) = DiffusionPrior(H, H_m)$
            \STATE  \\
            \COMMENT{training diffusion prior}\\
            
            \WHILE {true}
                \STATE $\h_b^{'}, \h_t \sim p(\h_b^{'}, \h_t | \h_m) $
                \STATE $\e_{t} \gets Quantize(\h_{t})$
                \STATE $\e_{b} \gets Quantize(\h_b^{'})$
                \STATE $\x = \D_b( Concat(\e_{t}, \e_{b}) )$
            \ENDWHILE 
            \STATE  \\
            \COMMENT{sampling procedure}\\
        
	\end{algorithmic}
\end{algorithm}

\section{Architecture Details and Hyperparameters}
\label{sec:network}
%The hierarchical VQ-VAE consists of two levels of VQ-VAE. 
The network parameters of the hierarchical VQ-VAE and associated training parameters are detailed in Tab.\ref{tab:tatbl1}. 
The network architecture for the prior model and the associated parameters for the diffusion process are detailed in Tab.\ref{tab:tatbl2}.

\begin{table}[h]
  \centering
  \begin{tabular}{@{}lc@{}}
    \toprule
    Hyperparameter & Value (Bottom Layer, Top Layer) \\
    \midrule
    Block number & 2, 1 \\
    Downsample rate & 4, 2 \\
    Codebook size & 512, 128\\
    Feature dimensions & 512, 512 \\
    Motion FPS & 60 \\
    Sequence length & 512 \\
    Batch size & 64 \\
    Optimizer & Adam \\
    Learning rate & 1e-4 \\
    Training epoch & 1000 \\
    
    \bottomrule
  \end{tabular}
  \caption{Hierarchical VQ-VAE hyperparameters.}
  \label{tab:tatbl1}
\end{table}

\begin{table}[h]
  \centering
  \begin{tabular}{@{}lc@{}}
    \toprule
    Hyperparameter & Value (Transformer Encoder) \\
    \midrule
    Number layers & 8 \\
    Number heads & 8 \\
    Latent dimensions & 512 \\
    Feed forward size  & 1024 \\
    Sequence length  & 128 \\
    Input dimensions & 1536 \\
    Condition Dimensions & 4800 \\
    Batch size & 64 \\
    Noise schedule & cosine \\
    DDPM Steps & 1000 \\
    Optimizer & Adan \\
    Learning rate & 4e-4 \\
    Droppout & 0.1 \\
    Train epoch & 500 \\    
    \bottomrule
  \end{tabular}
  \caption{Difusion Prior hyperparameters.}
  \label{tab:tatbl2}
\end{table}

\section{Video Result}
We provide a video demo to show the results of our method, including:

\noindent
\textbf{Comparisons with Other Methods. }
We compare our method with EDGE \cite{tseng2023edge} and Bailando \cite{siyao2022bailando}, in addition with the ground truth visual results on the AIST++ \cite{li2021ai} test set. 

\begin{figure*}[htp]
    \centering
    \includegraphics[width=0.8\linewidth]{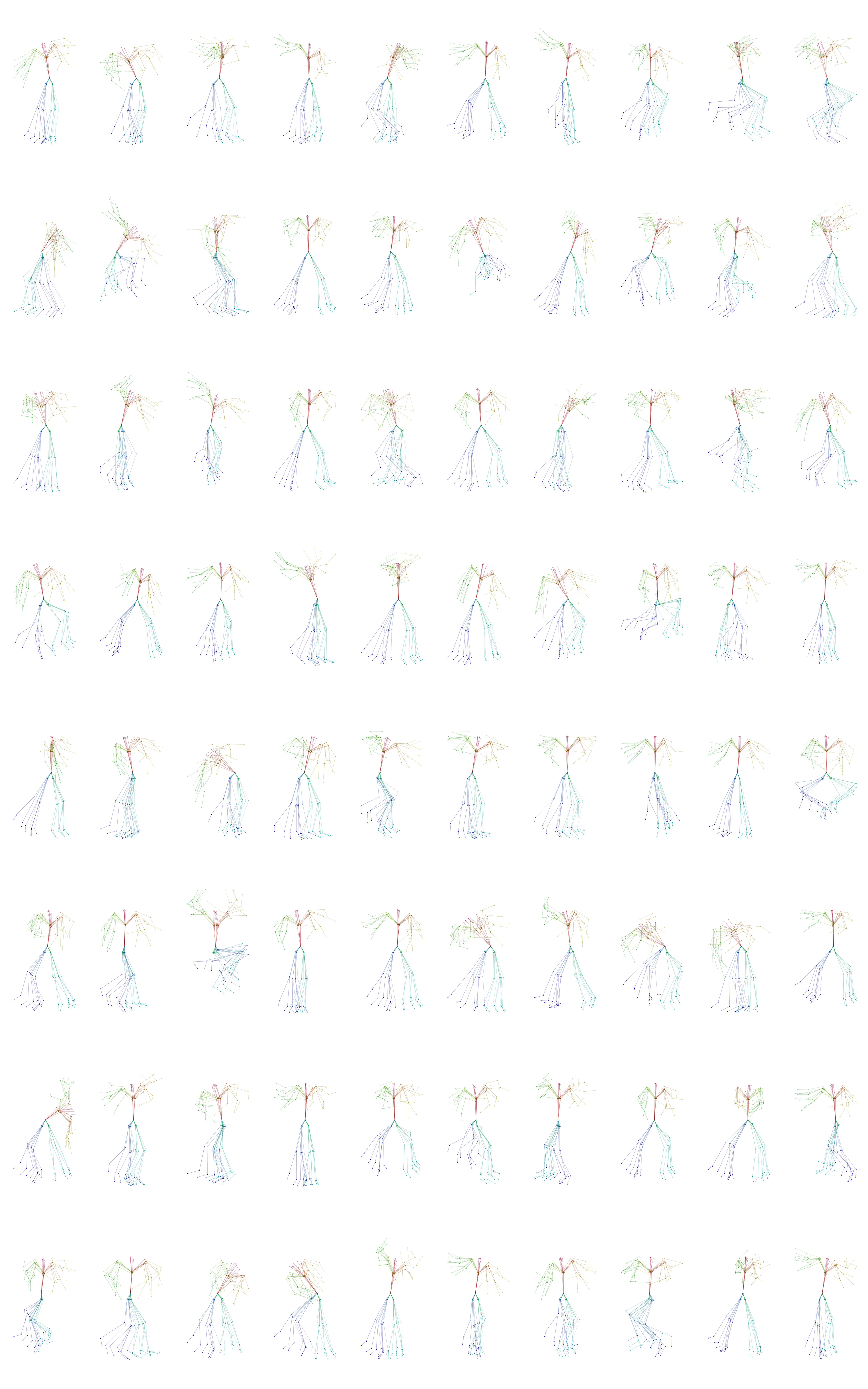}%\vspace{-0.2cm}
    \caption{More visual results of bottom codes. Each sample is the overlapped result of a specific bottom code combined with several random selected top codes.}
    \label{fig:fig_sm_bottom_analyse}%\vspace{-0.15cm}
\end{figure*}

\noindent
\textbf{the Interpretability of Latent Codes. }
More detailed visualizations of latent codes, including the interpretability of both bottom code and top code, as well as rendered results of swapping latent codes with a mixamo \cite{mixamo2023} character. Fig.\ref{fig:fig_sm_bottom_analyse} presents additional visualizations of the bottom code.
% For the interpretability of the bottom code, by fixing the top code and varying the bottom code, we can observed that different bottom codes correspond to different postures (dance poses).
% For the interpretability of the top code, we select a fixed bottom code and replace the bottom code corresponding to a segment of real dance motions with this fixed bottom code. From the synthesized results, it can be observed that the movements are constrained within a specific range, but the motion rhythm and trends remain consistent (dance movements).

% 视频demo展示对于latent codes更加详细的可视化结果，包括对于bottom code和top code的可解释性，以及交换latent codes的可视化结果。
% 对于bottom code的可解释性，通过固定top code而变换bottom code，视频中展示了不同的bottom code对应不同的posture。
% 对于top code的可解释性，我们首先选择一个固定的bottom code，然后把一段真实舞蹈动作对应的bottom code替换成该固定的bottom code，从合成的结果可以看到，动作被固定在一个限制的范围内，但是两者的运动节奏和运动趋势保持一致。

\noindent
\textbf{In the Wild Music Results. }
More rendered results on some in the wild music. %, including: Rechazame, Nolimit, TrainWreck and Yukiakari.

% \section{The Interpretability of Latent Codes}

% Fig.\ref{fig:fig_sm_bottom_analyse} presents additional visualizations of the bottom code. Please refer to the video demo for a more intuitive exploration of the visual results of the bottom code and top code.

% \begin{figure*}[htp]
%     \centering
%     \includegraphics[width=0.8\linewidth]{fig/sm_figure2.pdf}%\vspace{-0.2cm}
%     \caption{In the wild music results. Please refer to SM\_WILD\_MUSIC\_RESULTS.mp4.}
%     \label{fig:fig_sm_bottom_analyse}%\vspace{-0.15cm}
% \end{figure*}

% \section{In the Wild Music Results}
% In the wild music results. Please refer to SM\_WILD\_MUSIC\_RESULTS.mp4.

% \section{Applications}

% \subsection{Change bottom and top codes}

% \subsection{Editing}

%%%%%%%%% REFERENCES
{
    \small
    \bibliographystyle{ieeenat_fullname}
    \bibliography{main}
}